\documentclass[a4paper,11pt]{article}
\pdfoutput=1 % if your are submitting a pdflatex (i.e. if you have
\usepackage{jcappub} % for details on the use of the package, please
\usepackage[T1]{fontenc} % if needed
\newcommand{\R}{{\cal R}}
\newcommand{\Z}{{\cal Z}_0}
\newcommand{\M}{{\cal M}}

\newcommand{\bM}{{\bf M}}

\def\a{{\alpha}}
\def\b{{\beta}}
\def\l{{\lambda}}
\def\g{{\gamma}}
\usepackage{amsthm}
\usepackage{amsfonts}
\newtheorem{definition}{Definition}
\newtheorem{theorem1}{Theorem}
\newtheorem{proposition}{Proposition}
\usepackage{amsmath}
\usepackage{graphicx}

\title{\boldmath Asymptotic solution for expanding universe with matter-dominated evolution}

\author[1]{\v Z. Mijajlovi\'c,\note{Corresponding author.}}
\author{N. Pejovi\'c,}
\author{and V. Radovi\'c}

\affiliation{Faculty of Mathematics, University of Belgrade, Studentski trg 16, 11000 Belgrade, Serbia}

\emailAdd{zarkom@matf.bg.ac.rs}
\emailAdd{nada@matf.bg.ac.rs}
\emailAdd{rviktor@matf.bg.ac.rs}

\abstract{We applied the theory of regularly varying functions to
the analysis of  the cosmological parameters
for the $\Lambda$CDM model with the matter dominated evolution.
Carroll et al. proved in 1992 that for this type of universe with the curvature $k=0, -1$,
the expression $H(t)t$\, ($H(t)$ is the Hubble parameter) depends solely on the density parameter $\Omega(t)$.
Using this result and the theory of regular variation we infer
for such universe the complete asymptotics of all main cosmological parameters. More specifically, the following is derived.
If the limit  $\omega= \lim_{t\to\infty} \Omega(t)$ does exist and $\omega \not= 0$ then
the cosmological constant $\Lambda$ is equal to $0$.
If $\omega=0$ then for the expansion scale factor $a(t)$ we have $a(t)\sim e^{\sqrt{\Lambda/3}}$.
On the other hand, if the limit  $\lim_{t\to\infty} \Omega(t)$ does
not exist then   $a(t)$    bounces between
two power functions and therefore has infinitely many flexion points.
Hence, the deceleration parameter in this case changes the sign infinitely many times.}

\begin{document}
\flushbottom
\maketitle

\section{Introduction}

In our papers \citet{mijajlo2012} and \citet{mijajlo2015}, we applied the theory of regularly varying functions in the
asymptotic analysis of cosmological parameters of the expanding universe.
The main aim of the present paper is to apply this technique in  the study of the $\Lambda$CDM model with
a matter dominated evolution. As a result we obtained the complete asymptotics of
solutions of Friedmann equations and the related cosmological parameters. Our secondary
goal is to present further techniques of the theory of regular variation as a natural
method in the theoretical studies in cosmology.

The paper is organized as follows. In the introduction
we present some basic facts on Friedmann equations, short history and elements of regular
variation needed in the rest of the paper.
Also, properties of the constant $\Gamma$ introduced in \citet{mijajlo2012} are reviewed.
This constant and the related operator $\bf M$ will have to play the crucial role in  our
analysis of asymptotics of cosmological parameters. Due to the variety of appearance
in the literature,  in the section {\bf 2}
notation and the meaning od cosmological parameters are fixed. The section {\bf 3}
is central in this paper. There we inferred complete asymptotics of cosmological parameters
for a matter dominated $\Lambda$CDM model. By complete inference we mean that
we resolved the asymptotics when the limit $\lim_{t\to\infty} \Omega(t)$ exists,
but in the opposite case, too. This analysis is based on Carroll, Press and Turner
formulas for $\Omega(t)$ \citep{carroll} for this type of universe.
In the last section we demonstrated a use of the theory of regular variation
in possible descriptions of cosmological parameters of the dual universe.

We remind that the cosmological parameters are solutions of Friedmann equations:
\medskip

\begin{equation}\label{feq}
\begin{array}{ll}\displaystyle
\smallskip
\left(\frac{\dot{R}}{R}\right)^2= \frac{8\pi G}{3}\rho  -\frac{kc^2}{R^2}, &\quad \textrm{Friedmann equation},      \\
\smallskip
\displaystyle
\frac{\ddot{R}}{R}= -\frac{4\pi G}{3}\left(\rho + \frac{3p}{c^2}\right),   &\quad \textrm{Acceleration equation},   \\
\displaystyle
\dot\rho + 3\frac{\dot{R}}{R} \left(\rho + \frac{p}{c^2}\right)= 0,        &\quad \textrm{Fluid equation}.
\end{array}
\end{equation}
\medskip

Functions appearing in these equations are the expansion scale factor $R= R(t)$,
the energy density $\rho= \rho(t)$   and  the pressure of the material in the universe $p= p(t)$.
We show that all these parameters, including the Hubble parameter $H= H(t)= \dot R/R$   and
deceleration parameter $q= q(t)= - R\ddot R/ \dot R^2$, are regularly varying functions under standard circumstances.
The scale factor $R$ is often normalized in respect to the present epoch by $R= R_0 a(t)$, where $R_0= R(t_0)$ and $t_0$ is
a certain fixed time moment at the present.  If it is assumed that $R(t)$ is slowly varying in the present epoch
then $a(t)$ is the normalized scale factor having the value one at the present.
Observe that then the first Friedmann equation in $a(t)$ looks like:
\begin{equation}
\left(\frac{\dot{a}}{a}\right)^2= \frac{8\pi G}{3}\rho  -\frac{kc^2}{R_0^2a^2},
\end{equation}
while the fluid equation and the acceleration equation are invariant under the substitution $R(t)\rightarrow a(t)$.
In most cases we shall refer to $a(t)$ instead of $R(t)$.

The system (\ref{feq}) involves three unknown functions,
but only two of them are independent, for example the first and the third equation.
However the acceleration equation is  central  in our study of asymptotics of solutions of
Friedmann equations  for several reasons.
For instance it does not depend from the curvature index $k$. Further,
the theory of regularly varying solutions of such type of equations
can be applied successfully in the analysis of Friedmann equations even if the cosmological constant $\Lambda$ is added:
\begin{equation}
\begin{aligned}\label{AQL}
\left(\frac{\dot{a}}{a}\right)^2&= \frac{8\pi G}{3}\rho  -\frac{kc^2}{R_0^2 a^2} + \frac{\Lambda}{3},\\
\frac{\ddot{a}}{a}&= -\frac{4\pi G}{3}\left(\rho + \frac{3p}{c^2}\right) + \frac{\Lambda}{3}.
\end{aligned}
\end{equation}

\noindent
Namely, under the transformations
$ \rho'= \rho + \Lambda/(8\pi G), \enskip
p'  =  p   - \Lambda/(8\pi G) $
the Friedmann equations  (\ref{feq}) are invariant, while the fluid equation is not affected by the parameter $\Lambda$.

Theory of regularly varying functions is introduced mainly for studying behaviour of real functions at
infinity and also the functions satisfying the power law.
This theory was started in the thirties of the last century by J. Karamata in \citet{karamata}.
Many other authors continued to develop it \citep{bingham,seneta}. At the present time, this theory is used
in many areas of mathematics particularly in the asymptotic analysis of functions and probability theory.
There were  also some uses of this theory in
cosmology, particularly in the study of asymptotic behaviour of cosmological parameters, e.g. \citet{mijajlo2012,mijajlo2015}, but also by  \citet{molchanov,stern}.
\citet{barrow1996} and \citet{barrow2008} had a similar approach
in studies of asymptotic behaviour of solutions  to the Einstein equations describing expanding universes.
They used  there a theory of Hardy and Fowler which preceded the theory of regular variation.
We give a short review of main notions related to the regular variation and some extensions of this theory that
we shall need in the rest of the paper.

\subsection{Regularly varying functions}

A real positive continuous function
$L(t)$ defined for $x>x_0$  which satisfies
\begin{equation}\label{KI}
\frac{L(\lambda t)}{L(t)}\to 1\quad  {\rm as}\quad t\to \infty, \quad
\textstyle{\rm for \hskip 1mm  each \hskip 1mm real}\hskip 1mm \lambda>0,
\end{equation}
is called a slowly varying (SV) function.
Continuing  works of G.H. Hardy, J.L. Littlewood and  E. Landau,
Karamata \citet{karamata} originally defined and studied this notion for continuous functions.
Later  this theory was extended to measurable functions.
Due to physical constraints,  we assume here that all functions are continuously differentiable.

\begin{definition}
	A function F(t) is said to satisfy the  generalized power law if
	\begin{equation}\label{GPL}
	F(t)= t^rL(t)
	\end{equation}
	where $L(t)$ is a slowly varying function and $r$ is a real constant.
\end{definition}

Logarithmic function $\ln(x)$ and iterated logarithmic functions $\ln(\ldots\ln(x)\ldots)$ are examples of
slowly varying functions.
More complicated examples are provided in \citet{bingham,seneta} and \citet{maric}.

A positive continuous function $F$ defined for $t>t_0$,
is a regularly varying (RV) function of the index $r$,  if and only if it satisfies
\begin{equation}\label{KII}
\frac{F(\lambda t)}{F(t)}\to {\lambda}^r \quad {\rm as}\quad t\to \infty,\quad
\textstyle{\rm for \hskip 1mm  each}\hskip 1mm \lambda>0.
\end{equation}
It immediately follows that a regularly varying function $F(t)$ has the
form (\ref{GPL}).  Therefore $F(t)$
is regularly varying if and only if it  satisfies the generalized power law.
By $\cal R_\alpha$ we denote the class of regularly varying functions of index $\alpha$.
Hence ${\cal R}_0$ is the class of all slowly varying functions.
By ${\cal Z}_0$ we shall denote the class of zero functions at $\infty$, i.e.
$\varepsilon\in {\cal Z}_0$ if and only if $\displaystyle\lim_{t\to +\infty} \varepsilon(t)=0$.
The following representation theorem {\rm\citep{karamata} describes the  fundamental property of these  functions.
	\vskip 2mm
	\begin{theorem1}\label{KFT1}  {\it Representation theorem} $L\in\R_0$  is slowly varying function
	 if and only if there are measurable
		functions $h(x)$, $\varepsilon  \in \Z$ and $b \in \mathbb{R}$ so that
		\begin{equation}\label{RVrepresentation}
		L(x)= h(x) e^{\int_b^x\frac{\varepsilon(t)}{t}dt}, \quad x\geq b,
		\end{equation}
		and $ h(x)\to h_0$ as $x\to\infty$, $h_0$ is a positive constant.
	\end{theorem1}
	
	If $h(x)$ is a constant function,  then $L(x)$ is called normalized. A regularly varying function
	having normalized slowly varying part will be also called normalized. Let $\cal N$ denote
	the class of normalized slowly varying functions.
	The next  fact on $\cal N$-functions will be useful for our later discussion.
	If $L\in \cal N$ and there is $\ddot L$,
	then $\varepsilon$ in (\ref{RVrepresentation}) has the first order derivative $\dot\varepsilon$.
	This follows from the identity $\varepsilon(t)= t\dot L(t)/L(t)$.
	
	There are various classes of positive measurable functions with similar asymptotic behaviour
	to regularly varying function such as the class ER -- extended regularly varying
	functions (or Matuszewska class of functions),
	OR -- $O$-regularly varying functions, or recently introduced  \citep{cadena2014} interesting class $\mathcal{M}$.
	Here we shall particularly use the classes ER and OR. Extensive literature on these functions is available, e.g.
	\citet{aljancic,bingham,djurcic,seneta}.
	We review here, following  \citet{bingham}, their definitions and very  basic notions and properties related to these
	functions.   As in the case of regular variation  we
	assume that all mentioned functions are continuous.
	
	The limit in (\ref{KI}) does not exist always for an arbitrary function, but the limit superior and the limit inferior
	do exist. So let
	\begin{equation}\label{star}
	f^\ast(\lambda)= \limsup_{x\to \infty} \frac{f(\lambda x)}{f(x)},\quad f_\ast(\lambda)= \liminf_{x\to \infty} \frac{f(\lambda x)}{f(x)}
	\end{equation}
	Note that the difference $f^\ast(\lambda) - f_\ast(\lambda)$
	measures the oscillation of $f(\lambda x)/f(x)$ at infinity. It is said that
	a real positive measurable function $f$ belongs to the ER class if and only if there are constants
	$d$ and  $c$ such that
	\begin{equation}
	\lambda^d \leq f_\ast(\lambda) \leq f^\ast(\lambda) \leq \lambda^c,\quad \lambda\geq 1.
	\end{equation}
	The class OR functions is the set of positive measurable $f$ such that
	\begin{equation}
	0 < f_\ast(\lambda) \leq f^\ast(\lambda) < \infty,\quad \lambda\geq 1.
	\end{equation}
	We see at once that RV $\subseteq$ ER $\subseteq$ OR. There are examples that show that both extensions are proper.
	Note that $f$ is RV if and only if $f_\ast(\lambda) = f^\ast(\lambda)$, $\lambda\geq 1$.
	Therefore, RV functions are exactly OR (or ER) functions which do not oscillate at infinity.
	\medskip
	
	\begin{proposition}
		The following statements are equivalent for a real  function $f$:
		\begin{enumerate}
			\item $f$ belongs to the class ER.
			\item $f$ has the representation: $f(x)= \exp(C+ \eta(x) + \int_1^x \xi(t)dt/t)$,\quad $x\geq 1$,
			with $C$ constant, $\eta(x)\to 0$ as $x\to\infty$, $\xi$ bounded and $\xi$, $\eta$ both measurable.
		\end{enumerate}
	\end{proposition}
	\noindent
	Similar representation theorem holds for OR functions: $f\in$ OR if and only if
	\begin{equation}
	f(x)= \exp(\eta(x) + \int_1^x \xi(t)dt/t),\quad x\geq 1,
	\end{equation}
	where $\eta(x)$ and $\xi(x)$ are measurable and bounded.
	
	\subsection{The constant $\Gamma$}
	
	In our analysis an important role will have a constant $\Gamma$ which
	we introduced in \citet{mijajlo2012}. This constant  is related to the Friedmann equations,
	particularly to the acceleration equation
	which is considered there as a second order linear differential equation
	\begin{equation}\label{AQLM}
	\ddot a + \frac{\mu(t)}{t^2}a=0.
	\end{equation}
	Here
	\begin{equation}\label{mu2}
	\mu(t)=  \frac{4\pi G}{3}t^2\left(\bar \rho(t) + \frac{3\bar p(t)}{c^2}\right),
	\end{equation}
	where $\bar \rho(t)$ and $\bar p(t)$ are some particular solutions of Friedmann equations.
	Then $\Gamma$ is defined as the integral limit
	\begin{equation}\label{gamma}
	\Gamma = \Gamma_\mu = \displaystyle \lim_{x\to\infty} x\int_x^{\infty}\hskip -1mm \frac{\mu(t)}{t^2}dt \equiv \bM(\mu).
	\end{equation}
	assuming the existence of the limit integral on the right-hand side of (\ref{gamma}).
	The function $\bM(\mu)$ is a linear functional  $\bM\colon \M  \to R$,  where $\M$ is the space of
	all real functions satisfying the above integral condition (\ref{gamma}).
	$\M$ is called a Mari\'c class of functions (see \citet{mijajlo2012}).
	If this integral limit  does not exist, we say that $\Gamma$ does not exist.
	The constant $\Gamma$ appears in description of cosmological parameters by use of the theory of regularly varying
	solutions of linear second order differential equations, see \citet{mijajlo2012,mijajlo2015} and
	\citet{maric}.
	Determining the values of  $\Gamma$ one can obtain the asymptotical behaviour of the solutions
	$a(t)$, $\rho(t)$ and $p(t)$.
	The limit integral in (\ref{gamma}) in general is not easy to compute.  However, as
	\begin{equation}\label{WeakC}
	\lim_{t\to\infty}\mu(t)= \Gamma
	\end{equation}
	implies (\ref{gamma}),  we see that (\ref{WeakC})
	gives a  useful sufficient condition for the existence of regular solutions of the equation
	(\ref{AQLM}).
	%\vskip 2mm
	
	The significance of the integral limit (\ref{gamma}) is seen in the following facts.
	We have shown in \citet{mijajlo2012} that for any function $\mu\in \M$  such that
	$\Gamma_\mu < 1/4$, the Friedmann equations has a normalized regularly varying solutions and   the universe modelled by these solutions
	must be spatially flat or open.
	On the other hand, if $\Gamma_\mu > 1/4$ then the universe is oscillatory.
	Therefore, $\Gamma$  provides a kind of sharp threshold, or cut-off point,
	from which  the oscillation of $a(t)$ takes place \citep{hille}.
	It was also proved that  $\Gamma < 1/4$ implies a weak form of the equation of state.
	The converse also holds. Namely, we show:
	\begin{proposition} \label{prop2}
		For any normalized regularly varying solution of Friedman equations
		the integral limit  {\rm (\ref{gamma})} must exist and  $\Gamma \leq 1/4$.
	\end{proposition}
	
	{\bf Proof}.\, So suppose $a(t)= t^\alpha L$ is a solution of Friedmann equations where $L$ is a normalized slowly regular function and
	let $\beta= 1-\alpha$.  %and  $\Gamma= \alpha\beta$.
	In accordance with the representation (\ref{RVrepresentation}), Theorem 1, of slowly varying functions
	and as $L$ is normalized, we have $\dot L = \varepsilon L/t$, $\varepsilon\in {\cal Z}_0$.
	From there immediately follows the next derivation:
	\[
	\dot a  = \alpha t^{\alpha -1} L + t^\alpha \dot L= (\alpha + \varepsilon) t^{\alpha - 1} L, \phantom{\displaystyle\int_0^0}
	\]
	\[
	\ddot a = \dot\varepsilon t^{\alpha - 1}L - (\alpha +\varepsilon)(\beta - \varepsilon)t^{\alpha - 2}L,
\]
	wherefrom we obtain
	\vskip -4mm
	\[
	\displaystyle \frac{\ddot a}{a}= (\dot\varepsilon t  - (\alpha +\varepsilon)(\beta - \varepsilon))/t^2.
	\]
	As $a(t)$ is a solution of Friedmann equations, in particular of the equation (\ref{AQLM}), it must be
	$\mu= -\dot\varepsilon t  + (\alpha +\varepsilon)(\beta - \varepsilon)$
	and also
	\[
	\bM(\mu)= \bM( -\dot\varepsilon t  + (\alpha +\varepsilon)(\beta - \varepsilon)) =
	\]
	\[
	\bM( -\dot\varepsilon t) +  \bM((\alpha +\varepsilon)(\beta - \varepsilon)).
\]
	Further, $\bM( -\dot\varepsilon t)= 0$ (see proof of Theorem 2.2 in \citet{mijajlo2012}).
	Since $(\alpha +\varepsilon)(\beta - \varepsilon) \rightarrow \alpha\beta$ as $t\rightarrow \infty$, it
	follows $\bM((\alpha +\varepsilon)(\beta - \varepsilon))= \alpha\beta$. Hence
	$\bM(\mu)= \alpha\beta$ and so $\Gamma= \alpha\beta$, what proves the existence of $\Gamma$.
	Further, if $\alpha,\, \beta\geq 0$ then $\sqrt{\alpha\beta} \leq (\alpha + \beta)/2 = 1/2$, hence $\Gamma\leq 1/4$.
	If $\alpha < 0$ then $\beta > 0$, hence $\Gamma < 0 \leq 1/4$.
	\qed
	
	%Now we prove the inequality $\Gamma \leq 1/4$. Using the above derivation we see that for some $\xi \in {\cal Z}_0$
	%we have the following identities:
	%\begin{equation}\displaystyle
	%  t\frac{\dot a}{a}= \alpha + \varepsilon,\quad \left(t\frac{\dot a}{a}\right)^2= \alpha^2 + \xi
	%\end{equation}
	%Since $\ddot a/a= -\mu/t^2$ and by the obvious identity
	%$\ddot a/a = (\dot a/a)' + (\dot a/ a)^2$ it follows
	%\begin{equation}\displaystyle
	%  \left(\frac{\dot a}{a}\right)' + \left(\frac{\dot a}{a}\right)^2 + \frac{\mu}{t^2} = 0.
	%\end{equation}
	%Therefore, integrating the last identity and after multiplying by $x$ we obtain
	%\begin{equation}\displaystyle
	%  -x\frac{\dot a(x)}{a(x)} +  x\int_x^\infty \left(t \frac{\dot a(t)}{a(t)} \right)^2 t^{-2} dt + x\int_x^\infty \frac{\mu(t)}{t^2}dt = 0
	%\end{equation}
	%Finally, applying $x\to \infty$ to this identity we get $\alpha^2 - \alpha + \Gamma= 0$.
	%As $\alpha$ is a real number, for the discriminant $D= 1 - 4\Gamma$ it must be $D\geq 0$, i.e $\Gamma \leq 1/4$.
	
	Observe that $\alpha$ and $\beta$ are roots of the equation
	$x^2 - x + \Gamma= 0$ which defines the fundamental solutions of the acceleration equation, see \citet{maric,mijajlo2012}.

\section{Cosmological parameters and cosmological constants}

%{\bf Cosmological parameters}
%\medskip

In this section we review some  basic cosmological parameters and constants and their mostly well known properties.
Our only aim here is to establish their exact notation and meaning
used in the rest of the paper.

A cosmological parameter basically is a real function $P= P(t)$ of the time variable $t$ and it usually represents an
essential physical value related to the standard cosmological model. In the literature there are variations of their notation and
even definitions  that may lead to ambiguities. Therefore, we fix here their notation, at least for some of the parameters,  but we
assume their standard meaning. Also, for the simplicity of the notation from now on for the speed of light we shall take $c=1$.
%\vskip -5mm

\begin{equation}\label{param}
%\begin{align}
%\label{eqn:eqlabel}
\begin{array}{ll}
%\begin{split}
\smallskip
\rho_c= \rho_c(t)=  \displaystyle \frac{3H^2}{8\pi G}                    &\mbox{critical density},  \\
\smallskip

\Omega         = \Omega(t)         = \rho/\rho_c                           &\mbox{\rm density parameter},  \\
\smallskip

\Omega_\Lambda = \Omega_\Lambda(t) = \displaystyle\frac{\Lambda}{3H^2}     &\mbox{\rm $\Lambda$ density parameter},  \\
\smallskip

\Omega_k       = \Omega_k(t)       =\displaystyle  -\frac{k}{R^2H^2}     &\mbox{\rm curvature density parameter},   \\
\smallskip

q             =  q(t)             = \displaystyle -\frac{\ddot a}{aH^2}  &\mbox{deceleration parameter.}

%\end{split}
\end{array}
%\end{align}
\end{equation}

%{\bf Cosmological constants}
%\medskip

Derived constants are obtained from cosmological parameters fixing their values at $t=t_0$. The time value $t_0$
usually stands for a certain moment in the present epoch.
If $P$ is a cosmological parameter then we often write shortly $P_0$ instead of $P(t_0)$. Derived
constants inherit their names from the parameters from which they were obtained, e.g. Hubble constant $H_0$.
Here are some examples of derived  constants.

\begin{equation}\label{dcc}
\Omega_0= \frac{8\pi G}{3H_0^2}\rho_0,\quad \Omega_{\Lambda 0}=  \frac{\Lambda}{3H_0^2},\quad
\Omega_{k0}=  -\frac{k}{R_0^2H_0^2}
\end{equation}

The following form of Fiedmann equation is useful in determining
relations linking basic cosmological parameters
for pressureless universe \citep{carroll}.
It's usefulness follows from the explicit appearance of a term which relates the scale factor
and the density for this type of universe.
This theorem and it's proof also illustrate the use of derived cosmological constants as
the derived equation is in the parametrized form.
For the simplicity of the computation we took
the speed of light for unit, hence $c=1$.
\begin{theorem1}\label{FFE}
	The first Friedmann equation with non-zero cosmological constant $\Lambda$ is equivalent to the following equation:
\end{theorem1}
\vspace{-0.7cm}
\begin{equation}
	\begin{array}{ll}
&\dot a^2 + \frac{8\pi G}{3}(\rho_0/a^3 - \rho)a^2  = \\
& H_0^2(1 + \Omega_0(1/a - 1) + \Omega_{\Lambda 0}(a^2 - 1)).
\end{array}
\end{equation}
%\smallskip

{\bf Proof}
If $L= \displaystyle \dot a^2 + \frac{8\pi G}{3}(\rho_0/a^3 - \rho)a^2$ then using (\ref{AQL}) and definitions (\ref{dcc}) of
$\Omega_0$ and $\Omega_{\Lambda 0}$ we have
\begin{align*}
\smallskip
L &= \displaystyle
a^2\left(\frac{8\pi G}{3}\cdot\frac{\rho_0}{a^3} + \frac{\dot a^2}{a^2} - \frac{8\pi G}{3}\rho \right)  \\
\smallskip
&= \displaystyle
a^2\left(\frac{8\pi G}{3}\cdot\frac{\rho_0}{a^3} - \frac{k}{R_0^2a^2} + \frac{\Lambda}{3}\right) \\
\smallskip
&= \displaystyle
\frac{H_0^2}{a} + \frac{1}{a}\left( \frac{8\pi G}{3}\rho_0 - H_0^2 \right) - \frac{k}{R_0^2} + \frac{\Lambda}{3}a^2  \\
\smallskip
&= \displaystyle
\frac{H_0^2}{a} + \frac{1}{a}\left( \frac{k}{R_0^2} - \frac{\Lambda}{3} \right) - \frac{k}{R_0^2} + \frac{\Lambda}{3}a^2  \\
\smallskip
&= \displaystyle
H_0^2 + \left(H_0^2 + \frac{k}{R_0^2} - \frac{\Lambda}{3}\right)\left(\frac{1}{a} - 1\right) + \frac{\Lambda}{3}a^2 - \frac{\Lambda}{3}  \\
\smallskip
&= \displaystyle
H_0^2 + \frac{8\pi G}{3}\rho_0(1/a - 1) + \frac{\Lambda}{3}(a^2 - 1)  \\
\smallskip
&= \displaystyle
H_0^2\left(1 + \frac{8\pi G}{3H_0^2}\rho_0(1/a - 1) + \frac{\Lambda}{3H_0^2}(a^2 - 1) \right)  \\
\smallskip
&= \displaystyle
H_0^2(1 + \Omega_0(1/a - 1) + \Omega_{\Lambda 0}(a^2 - 1)).\quad\hfill \Box
\end{align*}

It is well known that a universe is pressureless (matter dominated) if and only if
$\rho= \rho_0/a^3$. Therefore, in this case the second term in $L$ vanishes and we have the
following statement.
%\smallskip

\noindent
{\bf Corollary} \citep[see ][]{carroll}\quad
If matter dominated universe is assumed, then the first Friedmann equation reduces to
the following equation with the constant coefficients:
\begin{equation}\label{crl}
\frac{{\dot a}^2}{H_0^2}= 1 + \Omega_0(1/a - 1) + \Omega_{\Lambda 0}(a^2 - 1)
\end{equation}

If a flat universe is assumed, i.e. $\Omega_k= 0$, due to the formula  $\Omega + \Omega_\Lambda + \Omega_k= 1$
the relation (\ref{crl}) obviously obtains more simple form:
\begin{equation}\label{cre}
\frac{\dot a^2}{H_0^2}= \Omega_0 a^{-1} + (1-\Omega_0)a^2.
\end{equation}

\noindent{\bf Note}.\,
By careful examination of the  proof of Theorem \ref{FFE}, we see that no infinitesimal transformation is applied
on $\Lambda$, i.e.  only algebraic transformations were used in this proof.
%It means that Theorem \ref{FFE} is also true if it is assumed that $\Lambda$ is a function on the time variable $t$.
We have the same conclusion for
formulas (\ref{crl}) and  (\ref{cre}).
%\smallskip

It is interesting that from (\ref{crl}) a rough estimation of $a(t)$ can be obtained in
regards to the various epochs in the development of the universe.
If the gravitation domination is assumed, i.e. the rate of the universe expansion is small
then the scale factor $a$ is small. Therefore the term $1/a$ dominates on the right-hand side
of the equation (\ref{crl}) and so $\dot a ^2 \propto a^{-1}$. From this relation
one obtains at once $a(t)\propto t^{\frac{2}{3}}$, a well known asymptotics for a matter
dominated universe. On the other side if the dark energy prevails, then the rate of the
universe expansion is large, i.e. now the term $a^2$ dominates the right-hand side of the
equation (\ref{crl}). Hence $\dot a^2/H_0^2 \sim\Omega_{\Lambda 0}a^2$, i.e.
$(\dot a/ a)^2 \sim \Lambda/3$,
wherefrom we obtain
$a(t)\sim \exp\left(\sqrt{\frac{\Lambda}{3}}t\right)$,
a well known result for this epoch of the universe evolution.

\section{Density parameter $\Omega(t)$ and regular variation}

As already noted the two Friedmann equations are not enough to fully solve for
the energy density,  the pressure and the scale factor. Hence we need an additional relation which connects
these parameters. Almost without exceptions equation of state $p= f(\rho)$ is assumed, usually
in the form $p = w \rho$, or  $p_i = w_i \rho$, if it assumed that the universe is composed from several
components $i$ (see page 6 in "The accelerating universe"\footnote{The Accelerating Universe, "The Nobel Prize in Physics 2011 - Advanced Information",   https://www.nobelprize.org/nobel\textunderscore prizes/physics/laureates/2011/ advanced-physicsprize2011.pdf}).
However, if the pressureless $\Lambda$CDM model is assumed, the regular variation property immanent
to the solutions of the system (\ref{feq}) and  to
other cosmological parameters as well,  enables us to use another approach.
Namely,  we can give without additional assumptions a complete asymptotics for this type of universe
using the theory of regularly varying functions. This section, including the subsections 3.1 and 3.2
are devoted to this analysis.
Hence, from now on we shall assume a pressureless universe with nonzero cosmological constant $\Lambda$.
For this type of universe  S.M. Carroll, W.H. Press and E.L. Turner %starting from (\ref{crl})
developed in \citet{carroll}  formulas of the form
\begin{equation}\label{cpt}
H(t)t= F(\Omega),\quad \mbox{\rm $\Omega= \Omega(t)$\,   is a density parameter},
\end{equation}
Our aim in this section is to show that under this assumption and from these relation
one can infer asymptotics for cosmological parameters regardless of the convergence of the limit integral (\ref{gamma}).
The functions $F(\Omega)$ are defined as follows.
\medskip

$F(\Omega)$ for flat Universe ($k= 0$):
\begin{equation} \label{Fk0}
F(\Omega)= \frac{2}{3}(1-\Omega)^{-\frac{1}{2}}\ln\left( \frac{1 + \sqrt{1-\Omega}}{\sqrt \Omega} \right)
\end{equation}

$F(\Omega)$ for open Universe ($k= -1$):
\begin{equation}
F(\Omega)= \frac{1}{1-\Omega}   -  \frac{\Omega}{2(1-\Omega)^{\frac{3}{2}}}\cosh^{-1}\left( \frac{2 - \Omega}{\Omega} \right)
\end{equation}
Observe that $F(\Omega)$ is a dimensionless physical quantity.

We note that the functions $F(\Omega)$ are easily inferred from (\ref{crl}).
Acceleration equation with $\Lambda$ is written as follows:
\begin{equation}\label{muL}
\frac{\ddot a}{a}= -\frac{\mu}{t^2} + \frac{\Lambda}{3}
\end{equation}

Taking in (\ref{mu2})
\begin{equation}
%\rho^\ast= \rho - \frac{\Lambda c^2}{8\pi G},\quad
%p^\ast   = p + \frac{\Lambda c^4}{8\pi G}.
\bar\rho = \rho^\ast - \frac{\Lambda }{8\pi G},\quad
\bar p   = p^\ast + \frac{\Lambda }{8\pi G}.
\end{equation}
%\begin{align*}
%  \rho_{\Lambda}&= \rho - \frac{\Lambda c^2}{8\pi G} \\
%  p_{\Lambda}  &= p+ \frac{\Lambda c^4}{8\pi G}
%\end{align*}
the acceleration equation is reduced to the standard form (\ref{feq}) where $\mu$ is replaced by
\begin{equation}\label{Lmut}
\mu_\Lambda=  (3\mu - \Lambda t^2)/3= 4\pi Gt^2(\rho^\ast + 3p^\ast)/3.
\end{equation}
In other words
\begin{equation}\label{muq}
\frac{\ddot a}{a}= -\frac{\mu_\Lambda}{t^2} \quad {\rm and}\quad  q= \frac{\mu_\Lambda}{F(\Omega)^2}.
\end{equation}

It is well known \citep[see ][and link https://ned.ipac.caltech.edu/level5/Carroll2]{liddle}} that in the  matter dominated
universe with the cosmological constant the following holds
\begin{equation}\label{qOL}
q= \Omega/2 - \Omega_\Lambda.
\end{equation}
Thus, for such universe,  by definition of $q(t)$, (\ref{muL})
and (\ref{qOL}) we have
\begin{equation*}
\Omega/2 - \Omega_\Lambda  = q= \frac{\mu}{(tH)^2} - \frac{\Lambda}{3H^2} = \frac{\mu}{F(\Omega)^2} - \Omega_\Lambda.
\end{equation*}

\begin{figure}[h]
	\centering
	\includegraphics[width=80mm,scale=1.0]{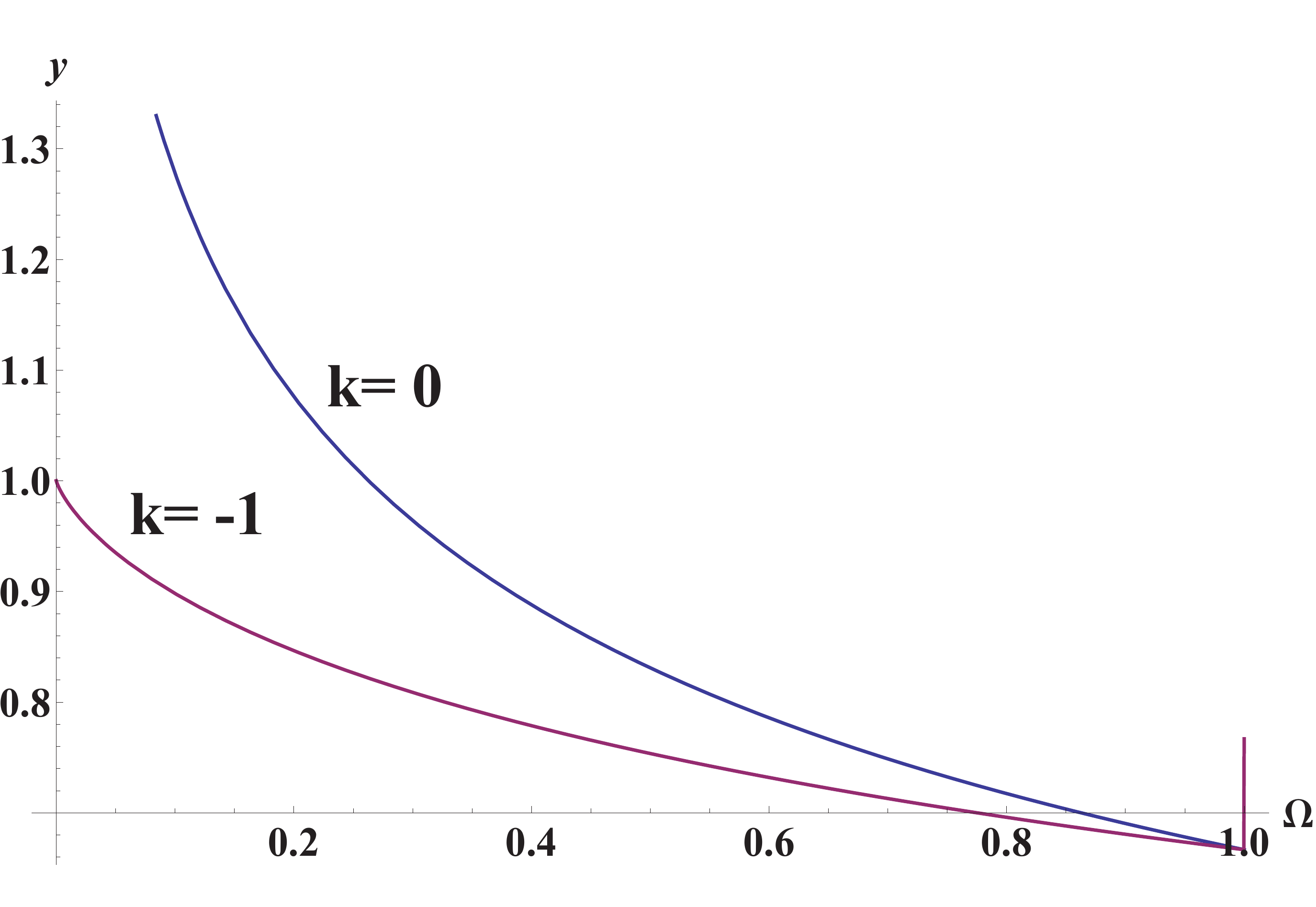}   %{ER01new.eps}
	\caption{Graphs of $y= F(\Omega)$}
	\label{Graph1}
\end{figure}
\begin{figure}[h]
	\centering
	\includegraphics[width=80mm, scale= 1.5]{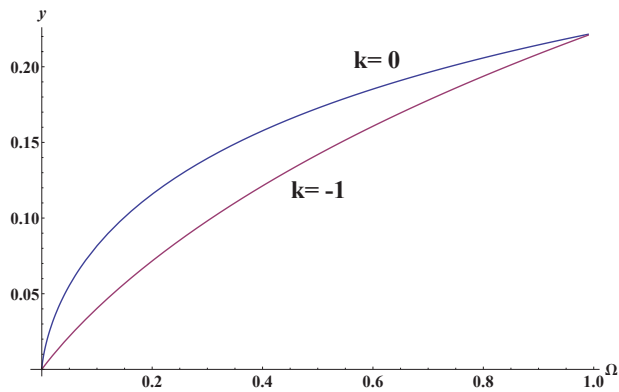}       %{ER02new.eps}  pdflatex: .png,  .pdf,  .jpg.
	\caption{Graphs of $\mu(\Omega)=   \displaystyle \frac{\Omega}{2}F(\Omega)^2$}
	\label{Graph2}
\end{figure}
Hence (see Figure \ref{Graph2})
\begin{equation}\label{muOFO}
\mu= \frac{\Omega}{2}F(\Omega)^2.
\end{equation}
We show how to infer asymptotic formulas for the scale factor $a(t)$
using $F(\Omega)$.
For this, we note that the Hubble parameter is the logarithmic derivative of $a(t)$:
\begin{equation*}
H(t) = \frac{\dot a(t)}{a(t)}= \frac{d\ln\, a(t)}{dt}.
\end{equation*}
Hence
\begin{equation}\label{HO}
\begin{array}{ll}
a(t)= &\exp\left(\int_{t_0}^t H(t) dt\right)= \exp\left(\int_{t_0}^t \frac{H(t)t}{t} dt\right)\\
=& \exp\left(\int_{t_0}^t \frac{F(\Omega)}{t} dt\right).
\end{array}
\end{equation}
We shall prove the following crucial characteristic of the scale factor $a(t)$:

\noindent
\begin{proposition}
	\begin{equation}\label{AER}
	\textrm{\it The function $a(t)$ belongs to the class {\rm ER}.}
	\end{equation}
\end{proposition}

\noindent{\bf Proof}\quad
This property will follow from (\ref{HO}) if we show that $F(\Omega(t))$ is bounded
on an interval $[t_0, \infty]$ for some $t_0 > 0$.
So first suppose that the universe is open. Then the function $F(\Omega(t))$ is obviously bounded, see Figure 1.
If the universe is flat, then $\Omega_\infty= 1$ and so  $F(\Omega(t))$ is bounded at the infinity, too.
Further, the value $\Omega_0= 0.3$ is widely taken  for the present epoch (see e.g. \citet{liddle}) and is close to the value preferred
by the observation. If we assume that the energy density $\rho$ becomes lower as the age of the universe becomes
older, we may suppose that the possible range for the constant $\Omega_\infty$ is the interval $[0.3,1]$ and so $F(\Omega(t))$
is bounded, too.
Therefore in all cases we may take that $F(\Omega(t))$ is bounded on the every interval $[t_0, \infty]$ for
every $t_0 > 0$. Hence, due to (\ref{HO}), is extended regularly varying function.
\quad\hfill $\Box$

In deriving asymptotic formula for  $a(t)$ and other cosmological parameters
we shall distinguish the following cases. The first one is when there is $\displaystyle\lim_{t\to\infty} \Omega(t)$.
The second one is when this limit does not exist.
\medskip

\subsection{There is a limit $\displaystyle\lim_{t\to\infty} \Omega(t)$}

In the greatest part of this section we shall discuss cosmological parameters for a pressureless universe
assuming that $\Omega(t)$ converges as $t\to \infty$ and for flat universe we shall obtain
their complete asymptotics. We shall briefly discuss asymptotics for an arbitrary universe, but not assuming 
the equation od state. 

Suppose first a pressureless universe and that there is a limit $\lim_{t\to\infty} \Omega(t)=\omega$.
From (\ref{Fk0}) we see that for the flat universe it must be $0\leq \omega \leq 1$. We shall distinguish
the cases $0< \omega \leq 1$ and $\omega=0$. Suppose first $0< \omega \leq 1$.
Then by the continuity of $F(\Omega)$,
there exists  $\lim_{\Omega\to \omega} F(\Omega)= \lim_{t\to \infty} F(\Omega(t))=  \alpha$ and by (\ref{cpt})
\begin{equation}\label{hubble}
H(t)= \frac{\alpha}{t} + \frac{\varepsilon}{t},\quad \varepsilon\in \Z.
\end{equation}
Then, using (\ref{HO}) it follows
\begin{equation}
\frac{a(t)}{a(t_0)}= \exp\left({\int_{t_0}^t \frac{\alpha}{t}dt + \int_{t_0}^t \frac{\varepsilon}{t} dt}\right)=
\left(\frac{t}{t_0}\right)^{\alpha} \exp\left({\int_{t_0}^t \frac{\varepsilon}{t}dt}\right).
\end{equation}
%\begin{align*}\label{HO}
%  \frac{a(t)}{a(t_0)} &= \exp\left({\int_{t_0}^t \frac{\Omega(t)}{t} dt}\right)  \\
%                      &= \exp\left(\int_{t_0}^t \frac{\alpha}{t} dt + \int_{t_0}^t \left( \frac{\Omega(t)- \alpha}{t} \right)dt\right).
%\end{align*}
Hence
\begin{equation}\label{apalpha}
a(t)= a(t_0) \left(\frac{t}{t_0}\right)^{\alpha}L_0(t), \quad t > t_0 >0.
\end{equation}
where $L_0(t)=  \exp(\int_{t_0}^t \frac{\varepsilon(t)}{t} dt)$, $\varepsilon(t) \rightarrow 0$ if $t\rightarrow\infty$.
Therefore $L_0(t)$ is slowly regular, and so $a(t)$ is a normalized regularly varying function.

Note that the  functions $F(\Omega)$ are defined for $0 < \Omega < 1$.
If $\Omega \sim 1$, then:
\medskip

\begin{equation*}
\begin{array}{rlrlrl}\hskip -8mm
\displaystyle
F(\Omega) &= \displaystyle\frac{2}{3}\left(1+ \frac{\sqrt{1-\Omega}}{1 + \sqrt{\Omega}}\right) + o(1-\Omega),           &&(k= 0).  \\
\phantom{F(\Omega)} &\phantom{= \frac{2}{3}\left(1+ \frac{\sqrt{1-\Omega}}{1 + \sqrt{\Omega}}\right) + o(1-\Omega),}    &&\phantom{(k= 0).}  \\
F(\Omega) &= \displaystyle\frac{2}{\Omega} - \frac{4}{3\Omega^2} + \frac{2\sqrt{1-\Omega}}{\Omega} + o(1-\Omega),\quad  &&(k= -1).
\end{array}
\end{equation*}
\smallskip

Hence, in both cases $\lim_{\Omega \to 1} F(\Omega) = 2/3$ as $\Omega\to 1$.
It is also easy to see that
\begin{equation*}
\begin{array}{rlrlrl}\hskip -8mm
F(\Omega) &\to +\infty \quad   &{\rm as}\quad   &\Omega\to 0^+\quad (k=0), \\
F(\Omega) &\to 1 \quad         &{\rm as}\quad   &\Omega\to 0^+\quad (k=-1).
\end{array}
\end{equation*}

Therefore, if $\omega= \lim_{t\to\infty} \Omega(t)$ exists where $\omega>0$  and $\lim_{t\to \infty} F(\Omega(t))=  \alpha$,
then the scale factor $a(t)$ satisfies generalized power law:
\begin{equation*}
a(t)= t^{\alpha} L_0(t),\quad L_0(t)\,\,\, {\rm is\,\, slowly\,\, varying}.
\end{equation*}
\noindent
According to \citet{mijajlo2012} then all cosmological parameters are uniquely determined and
a form of equation of state holds.

We shall consider in more details these parameters for the curvature index $k=0$.
Hence we assume until the end of this section $k=0$ and in this analysis we follow definitions (\ref{param}).
It will appear that the convergence of  $\Omega(t)$ to $\omega\not=0$ at infinity is a strong assumption.
It's consequence is that the cosmological constant is equal to 0.
To see this, first observe that from the identity $\Omega + \Omega_\Lambda =1$ it follows that there exists
$\omega_\Lambda= \lim_{t\to \infty} \Omega_\Lambda$. Hence
\begin{equation}\label{ool}
\omega + \omega_\Lambda= 1.
\end{equation}
As $\Lambda= 3\Omega_\Lambda H^2$, by (\ref{hubble}) there is a zero function $\xi= \xi(t)$ such that
\begin{equation}\label{Loa}
\Lambda= \frac{3(\omega_\lambda\alpha^2 + \xi)}{t^2}.
\end{equation}
Hence $\Lambda= 0$, $\omega_\Lambda=0$ and $\omega=1$ since
\begin{equation}\label{Lto0}
\frac{3(\omega_\lambda\alpha^2 + \xi)}{t^2} \to 0 \quad {\rm as}\quad t\to \infty.
\end{equation}

As $\lim_{\Omega\to 1} F(\Omega)= 2/3$, by (\ref{cpt}) we obtain $H(t)\sim \frac{2}{3}t$ and so $a(t)\sim (t/t_0)^{\frac{2}{3}}$
as $t\to +\infty$, as expected a well-known result for a matter dominated universe with $\Lambda=0$.

Now we shortly discuss asymptotics fo more interesting case, an arbitrary universe i.e. without assumptions of the matter dominance. 
However we assume $\Lambda=0$, but in this derivation we
do not assume equation of state.  In fact we shall obtain a weak form of it.
First we note that the threshold constant $\Gamma$ (see section 1.2) is in relation to acceleration equation and
it is defined by (\ref{gamma}). In  \citet{mijajlo2012} is shown that the exponent $\alpha$
of the expansion scale factor (\ref{apalpha}) is a solution of $x^2 - x + \Gamma=0$, hence
\begin{equation}\label{ga}
\Gamma= \alpha(1-\alpha).
\end{equation}

For the curvature index $k=0, -1$ it must be $\Gamma \leq 1/4$ (see Proposition (\ref{prop2}) and comments that precedes this proposition).
By (\ref{ga}) this condition is fulfilled since $\alpha \leq 2/3$.
Following ideas in \citet{mijajlo2012}, we define the equation of state constant $w$   by
\begin{equation}\label{wstate}
w= \frac{2}{3\alpha} - 1.
\end{equation}
Then the formulas for other cosmological parameters can be written as follows (see Theorem 3.5 in \citet{mijajlo2012}):
\begin{equation}\label{wparameters}
\begin{array}{rlrlrl}\hskip -8mm
\alpha&= \displaystyle\frac{2}{3(1+w)},                   &a(t)&= \displaystyle a_0t^{\frac{2}{3(1+w)}} L_0(t)    \\[10pt]
\hskip -3mm
%\rho  &\sim \displaystyle\frac{1}{6\pi G(1+w)^2t^2},    &p&\sim wc^2\rho                     \\[10pt]
H(t)  &\sim \displaystyle\frac{2}{3(1+w)t},\quad    &\bM(q)   &= \displaystyle\frac{1+3w}{2}
\end{array}
\end{equation}
We see that these formulas give the standard solutions of Friedman equations for the case $k=0$.
Also, there are functions $\hat w(t)$, $\xi(t)$ and $\zeta(t)$ such that
\begin{equation}\label{eqs}
p= \hat w \rho c^2,\quad ({\rm equation\,\, of\,\, state})
\end{equation}
where $\hat w(t)= w -t\dot\xi + \zeta$,\quad $\xi, \zeta \in \Z$.  Hence the weak form of the equation of state holds.
If $t\dot\xi \to 0$ as $t\to \infty$, then $\hat w(t)\approx w$,
what leads to $p= w\rho c^2$, the standard equation of state  and classical
asymptotics for cosmological parameters. In \citet{mijajlo2012} is also found
\begin{equation}\label{Gamma}
\bM(\mu)= \Gamma = \frac{2}{9} \cdot \frac{1+3w}{(1+w)^2}.
\end{equation}

However, there are several evidences measured in the last two decades that are against of 
$\Lambda= 0$. The first one is the estimated large value  $\Omega_\Lambda= 0.6911\pm 0.0062$,
according to results published by the Planck Collaboration in 2016 \citep{planck}, and the second one is the accelerated universe \citep{perlmutter,schmidt,riess}. Hence, it remains to
consider other two possibilities, $\omega=0$ and the divergent $\Omega$.
We shall discuss first
\smallskip

%	\bibitem{nobel2011} The Accelerating Universe, "The Nobel Prize in Physics 2011 - Advanced Information".

\noindent{\it Case} $\lim_{t\to\infty}\Omega(t)= 0$. Under this assumption and (\ref{Fk0}) we find
$F(\Omega) \sim \frac{2}{3}(\ln(2) - \frac{1}{2}\ln(\Omega))$ for $\Omega\to +0$, or
\begin{equation}\label{FO3}
F(\Omega)\sim -\frac{1}{3}\ln(\Omega)\quad {\rm as}\quad \Omega\to +0.
\end{equation}
Hence, by (\ref{cpt})  we have immediately
\begin{equation}
H(t)\sim -\frac{1}{3t}\ln(\Omega)\quad {\rm as}\quad t\to\infty .
\end{equation}

 As $H= \dot a/ a$ we also find for large $t$
\begin{equation}
a(t)= a(t_0) e^{-\int_{t_0}^t \frac{\ln(\Omega)}{3t}dt}.
\end{equation}
We can find more specific formula for the scale factor $a(t)$. By formulas (\ref{muOFO}) and (\ref{FO3}) we see that
$\mu\sim \Omega \ln(\Omega)^2/18$ as $\Omega\to +0$. Therefore (see also Figure \ref{Graph2})
\begin{equation}
\displaystyle \lim_{\Omega\to +0} \mu =0.
\end{equation}
Hence, the term $\mu/t^2$ may be neglected in the equation (\ref{muL}).
Assuming $\Lambda>0$, the fundamental solutions of so simplified equation
$\ddot a/a= \Lambda/3$ are $\exp(\sqrt{\Lambda/3}t)$ and $\exp(-\sqrt{\Lambda/3}t)$.
As the first fundamental solution dominates at infinity the second one, we find:
\begin{equation}
a(t) \sim  \exp(\sqrt{\Lambda/3}t)\quad {\rm as}\quad t\to +\infty.
\end{equation}
Interestingly, we obtained the same asymptotics for the scale factor $a(t)$ as in the case of cosmic inflation, perfectly consistent with
the ultimate fate of the universe such as Big Rip.

Now we turn to the next section where we consider the non-convergent $\Omega(t)$ at infinity.

\subsection{There is no limit $\displaystyle\lim_{t\to\infty} \Omega(t)$}

In this section we shall suppose that
the  limit $\lim_{t\to\infty} \Omega(t)$  does not exist.
However, by (\ref{AER}) $a(t)$ belongs to the class ER, hence we shall apply theory of these class
of function to the study of the asymptotic of $a(t)$.

The following simple statement will be useful in the next consideration.

\begin{proposition}
	Let $g(t)$, $f(t)$, $h(t)$ be real functions defined on a real interval $[t_0,\infty]$ and
	suppose they satisfy:
	\begin{equation*}
	g(t) \leq f(t) \leq h(t), \quad t\in [t_0,\infty].
	\end{equation*}
	Then there is a real function $u(t)$ such that
	\begin{equation}\label{fgh}
	f(t)= g(t)\cos(u(t))^2 + h(t)\sin(u(t))^2, \quad t\in [t_0,\infty].
	\end{equation}
	If $f$, $g$ and $h$ are continuously differentiable functions such that
	$g(t)<h(t)$, $t\in [t_0,\infty]$, then $u(t)$ is also continuously differentiable
	on $[t_0,\infty]$.
\end{proposition}
\noindent{\bf Proof}\quad
For any triple $a\leq x \leq b$ of real numbers there  are $\alpha, \beta \geq 0$ such that
$\alpha + \beta = 1$ and
$x= \alpha a + \beta b$. But then there is $u$ such that $\alpha= \cos(u)^2$ and hence
$\beta= \sin(u)^2$. Taking for each $t$, $a= g(t)$ and $b= h(t)$ we obtain (\ref{fgh}).
Further, it is easy to see that from (\ref{fgh}) follows
\begin{equation}\label{fgh2}
\displaystyle
f=  \frac{1}{2}(g + h) + \frac{1}{2}(g - h)\cos(2u)
\end{equation}
wherefrom
\begin{equation}
\displaystyle
u= \frac{1}{2}\arccos\left(\frac{h + g -2f}{h - g}\right),
\end{equation}
hence $u(t)$ is a continuously differentiable function.
\quad\hfill $\Box$

We have shown that $a(t)$ belongs to the class ER and that  $F(\Omega)$ is a bounded function for $t\geq t_0$ for $t_0> 0$.
Hence there are real numbers $\alpha$ and $\beta$ such that $\a < \b$ and $\alpha \leq \Omega(t) \leq \beta$.
By (\ref{AER}) and the representation  (\ref{HO})  we immediately infer
\begin{equation}
(t/t_0)^\a \leq a(t) \leq (t/t_0)^\b,\quad t\geq t_0.
\end{equation}
Then by the previous proposition there is a continually differential function  $u(t)$ such that
\begin{equation}\label{acs}
\displaystyle
a(t)= \frac{1}{2}\left(\left(\frac{t}{t_0}\right)^\a  + \left(\frac{t}{t_0}\right)^\b\right) +
\frac{1}{2}\left(\left(\frac{t}{t_0}\right)^\a  - \left(\frac{t}{t_0}\right)^\b\right)\cos(2u(t))
\end{equation}
or by normalizing, i.e putting $t_0=1$, we obtain somewhat simpler form
\begin{equation}\label{acossin}
\displaystyle
a(t)= \frac{1}{2}\left(t^\a  + t^\b\right) +
\frac{1}{2}\left(t^\a  - t^\b\right)\cos(2u)\equiv t^\a\cos(u)^2 + t^\b\sin(u)^2.
\end{equation}
\begin{figure}
	\centering
	\includegraphics[width=72mm, scale= 1.5]{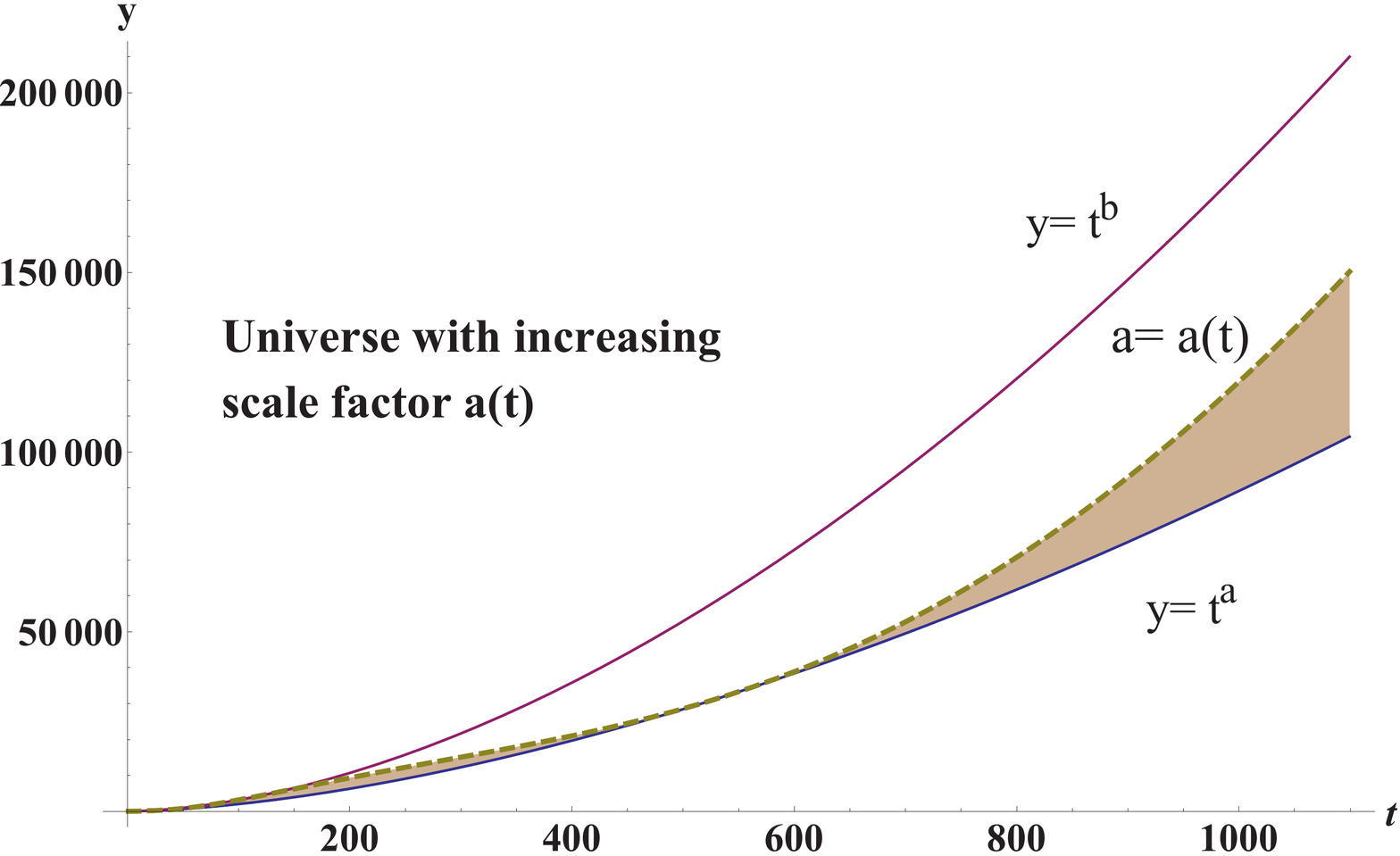} \includegraphics[width=72mm, scale= 1.5]{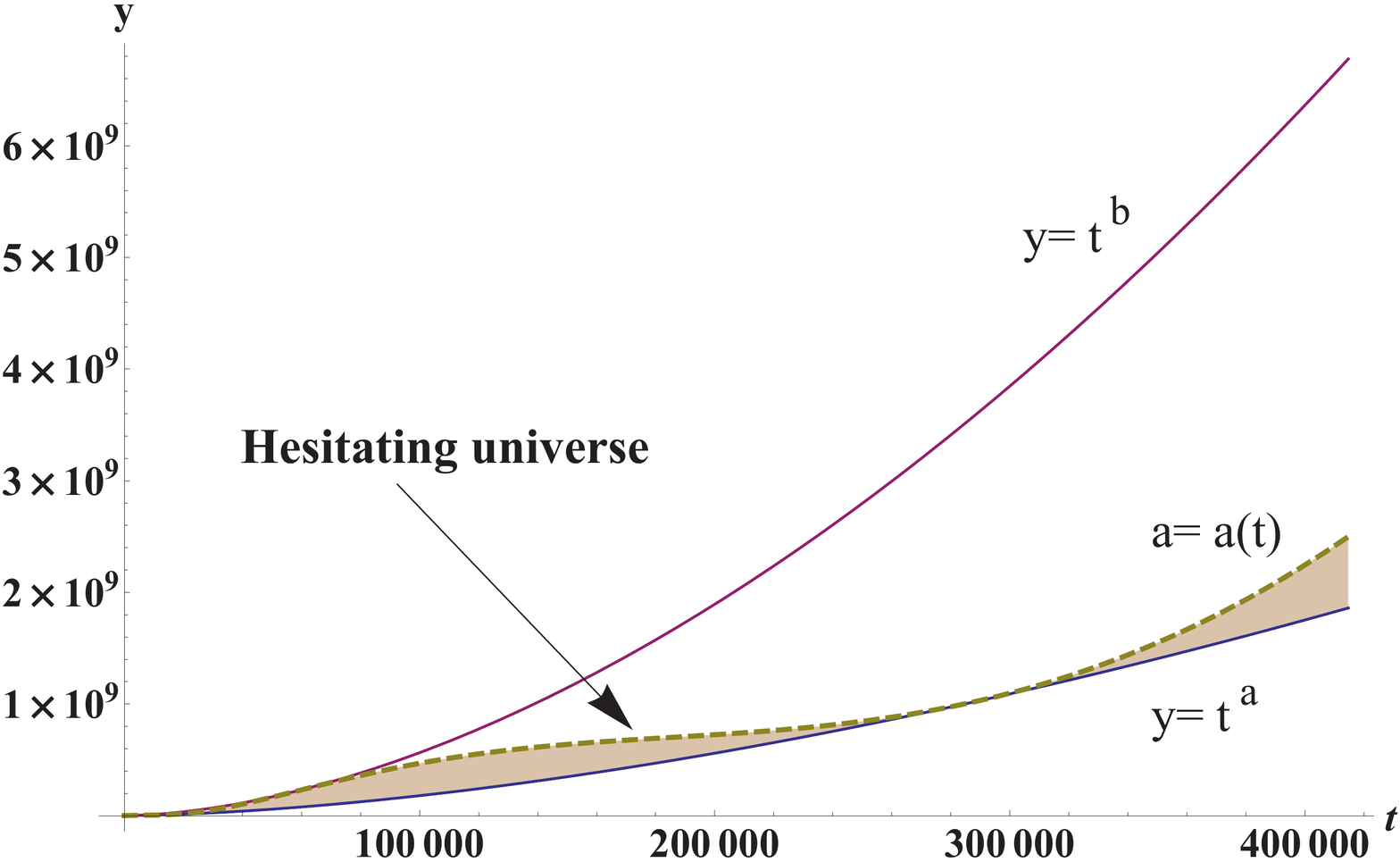}
	\caption{Evolution of the universe with increasing  (top) and hesitating (bottom) scale factor $a(t)$.} 
	\label{Graph3}
\end{figure}
\begin{figure}
	\centering
	\includegraphics[width=72mm, scale= 1.5]{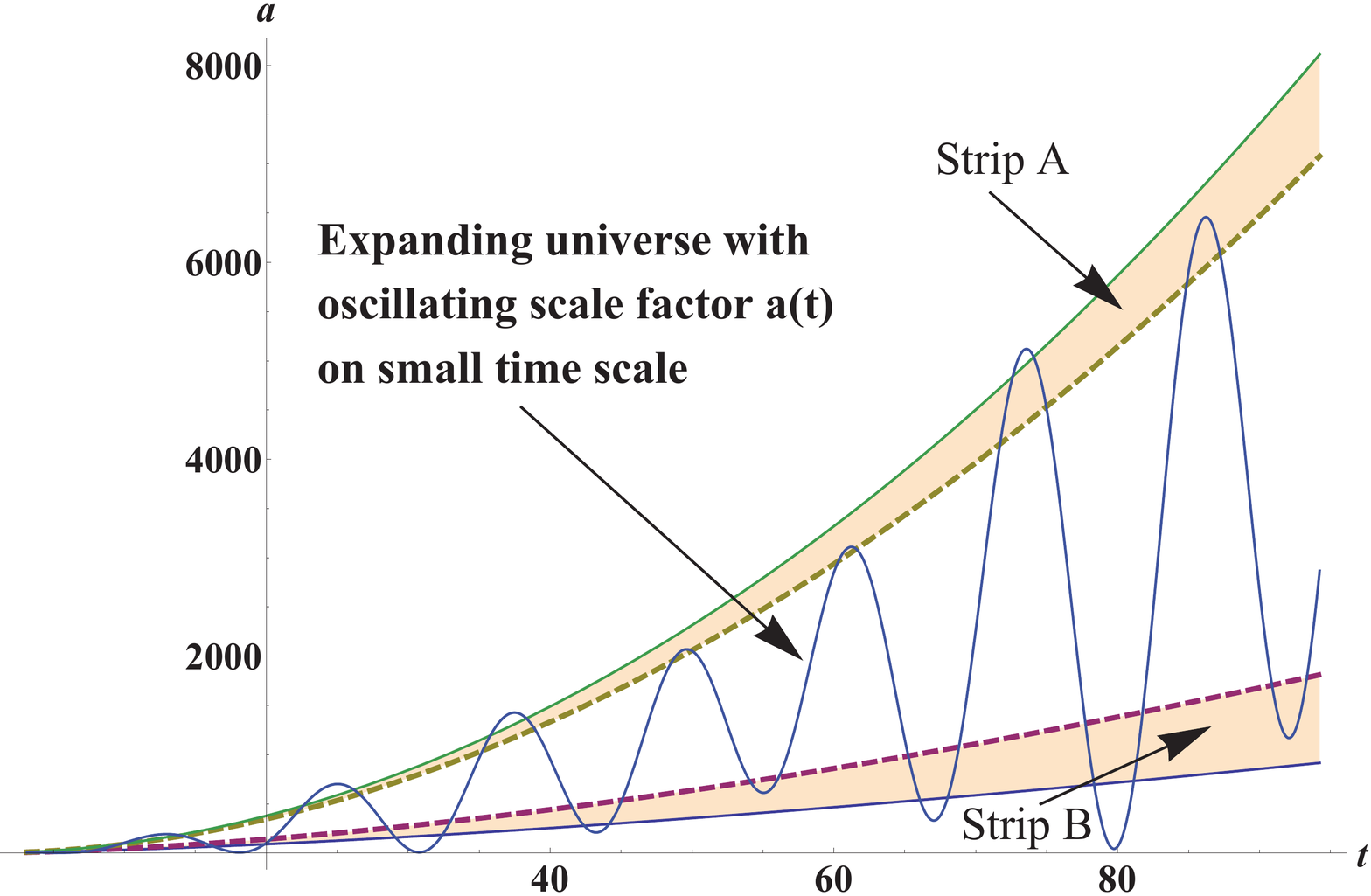} \includegraphics[width=72mm, scale= 1.5]{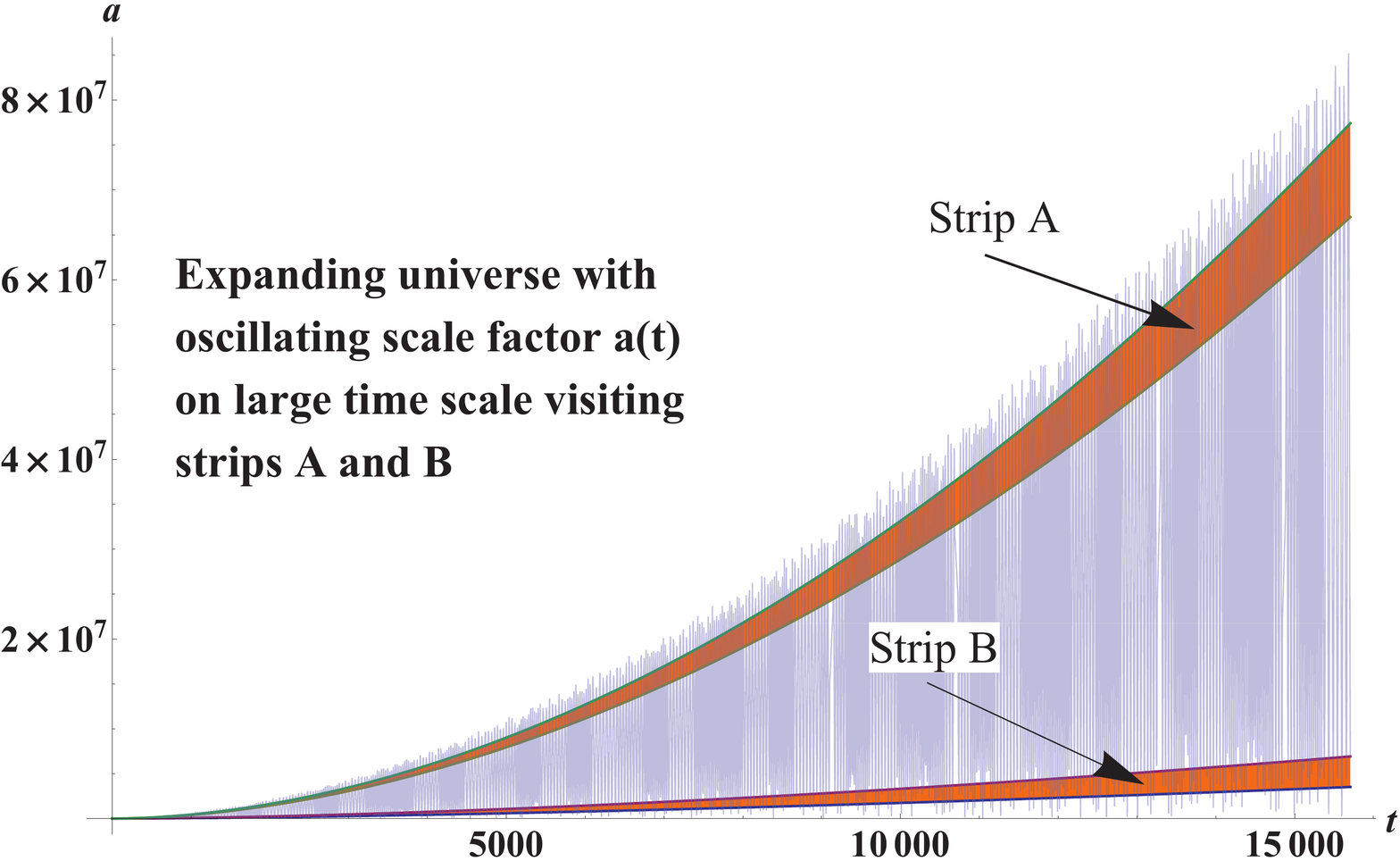}
	\caption{ Evolution of the oscillating scale factor $a(t)$ on small (top) and large (bottom) timescale.}
	\label{Graph4}
\end{figure}
Depending on the nature of the function $u(t)$, there are several possible cases in the evolution
of the scale factor $a(t)$.
These cases are depicted on the Figures 3. and 4.

If $\a\leq 0$ or $\b \leq 0$ then  the term in (\ref{acossin}) containing
the negative exponent may be omitted due to the largeness of $t$.
Therefore, in inferring above possibilities we suppose $0< \a < \b$. Then we have
\begin{equation}\label{dota}
\dot a= \a t^{\a -1}\cos(u)^2 + \b t^{\b -1}\sin(u)^2  +  (t^\b - t^\a )\dot u \sin(2u),
\end{equation}\label{ddota}
\vskip -11mm
\begin{multline}\displaystyle
\ddot a=  \frac{\a(\a-1)}{2} t^{\a-2}\cos(u)^2 + \frac{\b(\b-1)}{2} t^{\b-2}\sin(u)^2  +  \\
2(\b t^{\b - 1} - \a t^{\a - 1})\dot u \sin(2u) +
(t^\b - t^\a)(2\dot u \cos(2u) + \ddot u \sin(2u))
\end{multline}
Looking at (\ref{acs}) for large $t$ we can conclude:

\begin{proposition}\label{propa}
	\hspace{0.1cm}
\begin{itemize}
	\item[1.]
	The scale factor $a(t)$ bounces between bounding functions $t^\a$ and $t^\b$.
	\item[2.]
	If $\sin(u)\approx 0$, i.e. $u\approx k\pi$ for some integer $k$ then $a(t)\approx t^\a$.
	\item[3.]
	If $\sin(u) \not\approx 0$, then $a(t)\approx t^\a\sin(u)^2$.
	\item[4.]
	Every jump of $a(t)$ between $t^\a$ and $t^\b$ contains an inflection point, i.e. $a(t)$
	change from concave to convex, or vice versa. Therefore, $\ddot a$ changes the sign
	in the future infinitely many times and so does the deceleration parameter $q(t)$.
	Hence the universe's expansion alternatively  decelerates and accelerates.
\end{itemize}
\end{proposition}
Further we discuss the monotonicity of $a(t)$ at infinity (i.e. for large $t$). If in some point $t_1$,
$\dot u(t_1)= 0$, or $\sin(2u(t_1))= 0$,  then by ($\ref{dota})$ we have
$a(t_1)= \a t_1^{\a -1}\cos(u(t_1))^2 + \b t_1^{\beta -1}\sin(u(t_1))^2$, so
\begin{equation*}
\a t_1^{\a -1} \leq \dot a(t_1) \leq \b t_1^{\beta -1}.
\end{equation*}
Therefore $\dot a(t_1)> 0$ and $a(t)$ is  increasing  in a neighbourhood of $t_1$.

\noindent
If $\dot u(t) \sin(2u(t))\not\approx 0$, then
\begin{equation*}\label{das}
\dot a(t)\approx t^\b\dot u(t) \sin(2u(t)).
\end{equation*}
Suppose $u(t)$ is a time-like function, i.e. monotonously increasing and unbounded. Then
$\dot u(t) > 0$ and so the sign of $\dot a(t)$ depends solely on the term $\sin(2u(t))$.
As $u(t)$ is unbounded, $\sin(2u(t))$ certainly change the sign and so does $\dot a(t)$.
Hence $a(t)$ oscillates between $t^\a$ and $t^\b$ and the universe alternatively slows down and
increases the expansion.
But if $u(t)$ is a periodic function with the period of $\pi$, the same one as the function $\sin(2u)$,
But if $\dot u(t)$ and $\sin(2u(t))$ are functions of the same sign
then due to (\ref{dota}) for the sufficiently large $t$ we have
$\dot a(t)> 0$, i.e. $a(t)$ is increasing.

According to the previous discussion, all of the following cases are possible for the scale factor $a(t)$:
\begin{itemize}
	\item[1.] For large $t$, $a(t)$ is increasing.
	\item[2.] At some intervals $\dot a(t)\approx 0$, i.e. $a(t)$ is an almost  constant function on
	these intervals. This case corresponds to the so called Lemaitre hesitating universe.
	\item[3.] The scale factor $a(t)$ oscillatory varies between $t^\a$ and $t^\b$.
\end{itemize}
All these situations are depicted on the   diagrams of Figures 3. and  4. We also note that if
$2\dot u \cos(2u) + \ddot u \sin(2u)\not\approx 0$, then for large $t$
\begin{equation}
\ddot a\approx t^\b (2\dot u \cos(2u) + \ddot u \sin(2u)).
\end{equation}
The equation  (\ref{cpt}) tell us that $F(\Omega)= H(t)t$. Since $H= \dot a/a$, by (\ref{acossin}) and (\ref{dota})
taking $\g= \a -\b$ we have
\begin{equation}\displaystyle
F(\Omega)= \frac{\a t^\g \cos(u)^2 + \b \sin(u)^2 + t(1 - t^\g)\dot u \sin(2u)}{t^\g \cos(u)^2 + \sin(u)^2}
\end{equation}
For large $t$ we have $t^\g \approx 0$ as $\g < 0$. Therefore, if $u\not\approx k\pi$, $k$ is an integer, then we have
\begin{equation}\label{ouc}
F(\Omega)= \b + 2t\dot u \cot(u)
\end{equation}
On the other hand if $u\approx k\pi$, then we immediately find $F(\Omega)\approx \a$. Further, if we integrate
the equation (\ref{ouc}) we find
\begin{equation}
\ln|\sin(u)| = \frac{1}{2}\int_{t_0}^t \frac{F(\Omega)}{t} dt - \frac{1}{2}\b \ln(t),
\end{equation}
wherefrom and (\ref{HO}) we obtain $a(t)\approx t^\b \sin(u)^2$ for $u\not \approx k\pi$, an already noted
result  in Proposition \ref{propa}.3.

Finally let us mention the following useful result on ER functions. According to the theory of regularly
varying functions \citep{bingham} for an extended regularly varying function $f$, there are   numbers
$c'(f)$ and $d'(f)$, so called Karamata indices  and $\a'(f)$ and $\b'(f)$, so called  Matuszewska indices so that
 \citep[see ][, Theorem 2.1.8]{bingham}
\begin{equation}
{\l}^{d'(f)} \leq f_*(\l) \leq {\l}^{\b'(f)} \leq {\l}^{\a'(f)}\leq f^*(\l) \leq {\l}^{c'(f)},\quad \l\geq 1.
\end{equation}
Using this property and definition (\ref{star}) of functions $f_*$ and $f^*$ is is easy to show that there are numbers
$\a$, $\b$, $a$ and $b$ which naturally define strips $A$ and $B$ in the euclidean plane:
\begin{align*}
A&= \{(t,y)\in R^2\colon t\geq 0,\, t^b \leq y \leq t^\b \},\\
B&= \{(t,y)\in R^2\colon t\geq 0,\, t^a \leq y \leq t^\a \}.
\end{align*}
so that
\begin{equation}\label{strip}
f(t)\,\, \textrm{visits (intersects) strips $A$ and $B$ infinitely many times}.
\end{equation}
As the scale factor $a(t)$ belongs to the class ER, we see that $a(t)$ has the property
(\ref{strip}). This characteristic of $a(t)$ is depicted for a short time scale and a large time scale
on graphs, the Figure 4.

Even if the auxiliary function $u(t)$ in the previous analysis is unknown,
it has useful properties forced by
the ER property of $a(t)$ an the bounding functions $t^a$, $t^\alpha$, $t^b$ and $t^\beta$.
These properties were enough to deduce for divergent case of $\Omega(t)$, that $a(t)$ is oscillating
between the two indicated strips, has infinitely many flection points and that the deceleration
parameter $q(t)$ changes the sign infinitely  many times. These types of variation of the cosmological
parameters is difficult to explain by some inner evolution processes of the universe or the relicts from the
early universe as it was indicated in the convergent case. Possible explanation is given in the next section.

\section{Discussion and dual universe}

Here we shall discuss  certain  physical hypotheses for the universe models that might explain variation of
cosmological parameters we described in previous sections.
We briefly indicated some of these possibilities  in \citep{mijajlo2015}. For example,
one interesting possibility for the variations of  $q(t)$ and $p(t)$ for the convergent
$\Omega(t)$ could be a kind of reverberation due to
the extremely rapid expansion of the Universe  which appeared in the inflationary epoch,
about $10^{-36}$ seconds after the Big Bang. One can speculate that
these variations are the consequences of an echo effect due to thermalization which appeared when
the inflation epoch ended.
However a digressive variation for divergent $\Omega(t)$ of $a(t)$ and $q(t)$
on large time scale
is difficult to explain only with inner processes in the course of the universe evolution.
We prefer the idea that this variation is an effect of the existence of the dual universe
and  that it is a resultant of their mutual interference.
We shall discuss this possibility in more details.

One of the concepts of modern string theory and hence M-theory is that the big bang was a collision between two membranes 
\citep{khoury,jean,hong}.
The outcome was the creation of two universes, one in the surface of each membrane.
Using the Large Hadron Collider (LHC) located in CERN, some data are collected that might lead to the conclusion
that the parallel universe exist.
Specifically, if the LHC detects the presence of miniature black holes at certain energy levels,
then it is believed \citep{farag} that these would be the fingerprints of multiple universes.
Collected data are  still analyzed,
but there are also other research in this direction, e.g.  \citet{aguirre}.

We will not enter here into the full discussion on the existence of the multiverse.
But if it is assumed that the parallel universe exists, we can
explicitly find, at least for regularly varying cosmological parameters,
a set of formulas that might represent cosmological parameters of a dual universe.
Our argument for supporting this approach we found in the next theorem and the symmetry
which exists between formulas expressing cosmological parameters for this pair of universes.
The symmetry is represented by the Galois group of the associated algebraic equation (\ref{EQ144}).
This group and the translation of the set of formulas of the primary universe to it's dual
is also described in this section.
We obtain them using  the second fundamental solution $L_2(t)$ in Howard - Mari\'c
theorem \citep[see ][]{maric,mijajlo2012}) applied to the acceleration equation.
\vskip 2mm

\begin{theorem1}\label{MaricT} (Howard-Mari\'c)
	Let $-\infty< \Gamma <1/4$, and let $\alpha_1<\alpha_2$ be two
	roots of  the equation
	\begin{equation}\label{EQ144}
	x^2 - x +\Gamma= 0.
	\end{equation}
	Further let $L_i$, i=1,2 denote two normalized slowly varying functions.
	Then there are two linearly independent regularly varying solutions of\, $\ddot{y} + f(t)y=0$\, of the
	form
	\begin{equation}\label{EQ145}
	y_i(t) = t^{\alpha_i}L_i(t), \quad i=1,2,
	\end{equation}
	if and only if $\displaystyle \lim_{x\to\infty} x\int_x^{\infty}\hskip -2mm f(t)dt= \Gamma$.
	Moreover,  $\displaystyle L_2(t)\sim \frac{1}{(1-2\alpha_1)L_1(t)}$. \qed
	
\end{theorem1}
\vskip 2mm

The second fundamental solution and so  the dual
set of these formulas  is determined by the second root $\b= 1-\a$  of the
quadratic equation $x^2-x + \Gamma= 0$ appearing in this theorem.
To avoid singularities, we assume $\a, \b \not=0$.
Now we use $\beta$ instead of $\alpha$ for the index of RV
solution $a(t)$ - deceleration parameter and for determination of other
constants and cosmological parameters. As in (\ref{wstate}) we introduce
$w_\beta= \frac{2}{3\beta}-1$. Then we have
the following symmetric identity for equation of state parameters:
\begin{equation}\label{symmetricb}
w_\alpha + w_\beta+ 3w_\alpha w_\beta = 1
\end{equation}
For our universe we have $w= w_\alpha$, while for the dual
universe the corresponding equation of state parameter is $w_\beta$.
Then the dual formulas are obtained by replacing $\alpha$ with $\beta$ and$w_\alpha$ with $w_\beta$ in
(\ref{hubble}), (\ref{eqs})  and (\ref{wparameters}).
If one wants to give any
physical meaning to the so obtained dual set of functions, it is rather natural
to interpret them as the cosmological
parameters of the dual universe.

As we shall see these two universes are isomorphic in the sense
that there is an isomorphism which maps cosmological parameters into their dual forms.
Hence, this isomorphism introduces a symmetry between
cosmological parameters and their dual forms. It would mean that both universes have same or similar physics
as anticipated in \citet{hong}.
In this derivation we  use some elements of the Galois theory. For the basics of this
theory the reader may consult for example \citep{lang}.

Our assumption   $\Gamma<\frac{1}{4}$ and that the solutions
$\alpha$ and $\beta$ of the equation (\ref{EQ144}) differ, say $\alpha<\beta$, introduces
the following kind of symmetry. Let $F= {\bf R}(t,\Gamma)$ be the extension algebraic field where $\bf R$ is
the field of real numbers and $t$ and $\Gamma$ are letters (variables).
It is easy to see that for such $\Gamma$ the
polynomial $x^2-x+\Gamma$ is irreducible over the field $F$. Hence, the Galois group $\bf G$
of the equation (\ref{EQ144}) is of the order 2 and has a nontrivial automorphism $\sigma$.
Let $\alpha$ and $\beta$ be the roots of the polynomial $x^2-x+\Gamma$.
Then $\sigma(\alpha)=\beta$ and $\sigma(\beta)=\alpha$.
Further, let
$
\Gamma= \frac{2}{9}\cdot\frac{1+3w}{(1+w)^2}
$
where $w$ is a parameter. Then we can take $\alpha= \frac{2}{3(1+w)}$  and $\beta= \frac{1+3w}{3(1+w)}$.
Let $w_\alpha\equiv w$ and
$w_\beta\equiv \frac{1-w}{1+3w}$. Then
$\sigma(w_\alpha)= w_\beta$ since $w_\alpha$ and $w_\beta$ are rational expressions respectively in $\alpha$ and $\beta$.
Further, the time $t$ and the constant $\Gamma$ are invariant under $\sigma$ i.e. $\sigma(t)=t$
and $\sigma(\Gamma)= \Gamma$ since $t$ and $\Gamma$ are the elements of the ground field $F$.
The cosmological parameters (\ref{hubble}) and (\ref{wparameters})
are rational expressions of $w$  so if $P_\alpha$
is the corresponding parameter to the solution $\alpha$, then $\sigma(P_\alpha)= P_\beta$.
For example, for the Hubble  parameters we have $\sigma(H_\alpha)= H_\beta$.
Hence, not only solutions (isomorphic via $\sigma$) come into the pairs but
the sets of all cosmological parameters   come as well. At this point one may speculate about
two dual mutually interacting universes having the same time $t$ and the constant $\Gamma$ and the conjugated
parameters $w_\alpha$ and $w_\beta$ connected by the relation (\ref{symmetricb}).

Of course, there is a question what are the values of the constants appearing in cosmological parameters,
for example of $w= w_\alpha$.
During the last two decades \citep{peeble} there was a great progress in fundamental
measurements of cosmic data and estimation of cosmological constants.
In particular collected precision data in  WMAP\footnote{https://lambda.gsfc.nasa.gov/product/map/dr5/} (Wilkinson Microwave Anisotropy Probe)  mission.
enabled accurate testing of cosmological models. It was found that
the emerging standard model of cosmology, a flat $\Lambda$-dominated universe seeded by a
nearly scale-invariant adiabatic Gaussian fluctuations, fits the WMAP data \citep{spergel}.
According to these observations, the value of $w$, also called equation of state of cosmological constant
(or the dark energy equation of state),
is near $-1$. More precisely,  the collected data are consistent with the density being time-independent as for
a simple cosmological constant ($w = -1$), with uncertainties in $w$ at the 20\% level \citep{tegmark}.
Other results from experimental cosmology,
such as the Baryon Oscillation Spectroscopic Survey (BOSS) of Luminous Red Galaxies (LRGs) in
the Sloan Digital Sky Survey (SDSS) are also in the favor of  $w = -1$, \citep{anderson}, Particle Data Group: http://pdg.lbl.gov.
However, the value $w= -1$ yields singularity in (\ref{wparameters}).
For such $w$ there is no corresponding $\alpha$ neither $\Gamma$.
Equation of state is $p= -\rho c^2$ and then by fluid equation we have $\dot{\rho}= 0$, i.e $\rho$ is constant.
This case corresponds to the cosmological constant, so $\rho= \rho_\Lambda = \frac{\Lambda}{8\pi G}$.
In the absences of $\alpha$ and $\beta$ for dual $w_\beta$ of $w=w_\alpha$  we may take
(\ref{symmetricb})  for defining relation . Putting $w_\alpha= -1$ in this identity we obtain $w_\beta= -1$.
Hence, if the $\Lambda$CDM model is assumed, dual universe is
also equipped wit cosmological constant and its expansion is also governed by  dark energy.

The other values of $w$ are also considered.
For example if  $w=1/3$ then $\alpha= \beta= 1/2$, $\Gamma= 1/4$ and in this case
Howard-Mari\'c theorem cannot be applied since functions $L_1(t)$ and $L_2(t)$ from this theorem are not fundamental
solutions.  But there is a variant of this theorem appropriate for this case \citep{maric}, and applying it
one can show that $a(t)$ is regularly varying of index $\frac{1}{2}$
if and only if $w\sim \frac{1}{3}$ as $t\to \infty$, i.e.  $p\sim \frac{1}{3}c^2 \rho$
holds asymptotically. This is the second classic cosmological solution.
For more details one can consult  \citet{mijajlo2012}.

\section{Conclusion}

We analysed  Friedmann equations and cosmological parameters from the point of view of regular variation.
The central role in this analysis had the acceleration equation since
it can be considered as a linear second order differential equation and that the theory
of regularly varying solutions of such equations is well developed \citep{maric}.
Particular attention in our discussion was given to a pressureless universe with non-zero cosmological constant $\Lambda$.
For this type of universe  S.M. Carroll, W.H. Press and E.L. Turner
introduced in \citet{carroll} a function $F(\Omega)$ by which the other cosmological parameters can be expressed.
This allowed us, using the theory of regular variation, to describe complete asymptotics for cosmological parameters
when $\omega= \lim_{t\to\infty} \Omega$ does   exist, but also
in the case when this limit does not exist.
As a conclusion we obtained for the convergent $\Omega(t)$  at infinity and $\omega\not=0$
that  $\Lambda= 0$. We also found asymptotics for all main cosmological parameters if $\omega=0$.
On the other hand, for divergent $\Omega$ we got that the expansion scale factor $a(t)$ varies between two
strips bounded by power functions and that it changes it's convexity and concavity infinitely many times.
Hence, in this case the deceleration parameter changes it's sign infinitely many times and $a(t)$
must accelerate and decelerate infinitely many times, too.
There are strong evidences in favour of the existence of the dark energy.
For example  \citet{tegmark}, list three such observational proofs: supernovae  of type Ia,
power spectrum analysis  as done in  \citet{tegmark} and the late ISW  effect (Integrated Sachs--Wolfe effect).
Therefore if the $\Lambda$CDM model is assumed, the case $\omega\not=0$ is excluded.
We also discussed  the threshold constant $\Gamma$ and
the formally introduced equation of state parameter $w$. Both constants have an important role
in describing asymptotics of cosmological parameters and evolution of the Universe.
We also presented asymptotic formulas that might represent the cosmological parameters of
the interacting dual universe.

\acknowledgments
	This work was supported by the Serbian Ministry of Science, grant number III44006



\begin{thebibliography}{99}


\bibitem[\protect\citeauthoryear{Aguirre, Johnson \& Shomer}{2007}]{aguirre} Aguirre A., Johnson M.~C., Shomer A., {\it  Towards observable signatures of other bubble universes}, PhRvD, 76, (2007), 063509 



\bibitem[\protect\citeauthoryear{Aljan\v ci\'c \& Arandjelovi\'c}{1977}]{aljancic}  Aljan\v ci\'c, S.,  Arandjelovi\'c, D.,{\it $O$-regularly varying functions}, Pub. Inst. Math., 22(36), (1977), 5-22s


\bibitem[\protect\citeauthoryear{Ali, Faizal \& Khalil}{2015}]{farag} Ali A.~F., Faizal M., Khalil M.~M., {\it Absence of black holes at LHC due to gravity's rainbow}, PhLB, 743 (2015), 295 


\bibitem[\protect\citeauthoryear{Anderson et al.}{2014}]{anderson} Anderson L., et al., {\it The clustering of galaxies in the SDSS-III Baryon Oscillation Spectroscopic Survey: baryon acoustic oscillations in the Data Releases 10 and 11 Galaxy samples}, MNRAS, 441, (2014), 24 

\bibitem[\protect\citeauthoryear{Bingham, Goldie \& Teugels}{1987}]{bingham} Bingham, N.H., Goldie, C.M., Teugels, J.L., {\it Regular  variation}, Cambridge Univ. Press, Cambridge (1987)
	
\bibitem[\protect\citeauthoryear{Barrow}{1996}]{barrow1996} Barrow J.~D., {\it Varieties of expanding universe}, CQGra, 13 (1996), 2965 

\bibitem[\protect\citeauthoryear{Barrow \& Shaw}{2008}]{barrow2008} Barrow J.~D., Shaw D.~J., {\it Some late-time asymptotics of general scalar tensor cosmologies}, CQGra, 25 (2008), 085012 


\bibitem[\protect\citeauthoryear{Cadena \& Kratz}{2014}]{cadena2014} Cadena, M., Kratz, M., {\it An extension of the class of regularly varying functions}, https://hal-essec.archives-ouvertes.fr/hal-01097780 (2014), 35p
	
	
\bibitem[\protect\citeauthoryear{Carroll, Press \& Turner}{1992}]{carroll} Carroll S.~M., Press W.~H., Turner E.~L., {\it The cosmological constant}, ARA\&A, 30 (1992), 499 
	
\bibitem[\protect\citeauthoryear{Carroll}{2001}]{carroll2000} Carroll S.~M., {\it The Cosmological constant}, LRR, 4 (2001), 1 
	

\bibitem[\protect\citeauthoryear{Djur\v ci\'c}{1997}]{djurcic}  Djur\v ci\'c, D., {\it $O$-regularly varying functions and some asymptotic relations}, Pub. Inst. Math. 61 (1997), 44-52


	%\bibitem{coles} Coles, P. \& Lucchin, F., Cosmology:  the Origin and Evolution of Cosmic Structure (2nd ed.).
	%           Wiley, Chichester(2002)

\bibitem[\protect\citeauthoryear{Friedmann}{1924}]{friedmann} Friedmann A., {\it {\" U}ber die M{\" o}glichkeit einer Welt mit
              konstanter negativer Kr{\" u}mmung des Raumes}, ZPhy, 21 (1924), 326 

\bibitem[\protect\citeauthoryear{Hille}{1948}]{hille} Friedmann A., {\it Non-oscillation theorems}, Trans. Amer. Math. Soc., 64 (1948), 234 


\bibitem[\protect\citeauthoryear{Cai et al.}{2007}]{hong} Cai Y.-F., Li H., Piao Y.-S., Zhang X., {\it Cosmic duality in quintom Universe}, PhLB, 646 (2007), 141 

\bibitem[\protect\citeauthoryear{Howard \& Mari\'c}{1997}]{howard} Howard, H.C., Maric, V., {\it Regularity and nonoscillation of solutions of second order linear differential equations}, Bull. T. CXIV de Acad. Serbe Sci et Arts, Classe Sci. mat. nat, 22 (1997), 85-98
	
	%\bibitem{islam} Islam, J.N.: An introduction to mathematical cosmology. Cambridge Univ. Press, Cambridge (2004).
	
\bibitem[\protect\citeauthoryear{Karamata}{1930}]{karamata} Karamata, J, {\it Sur une mode de croissance r\'eguliere fonctions},  Math. (1930)

\bibitem[\protect\citeauthoryear{Khoury et al.}{2001}]{khoury} Khoury J., Ovrut B.~A., Steinhardt P.~J., Turok N., {\it The Ekpyrotic Universe: Colliding Branes and the Origin of the Hot Big Bang}, PhRvD, 64 (2001), 123522 

\bibitem[\protect\citeauthoryear{ Kusano \&  Mari\'c}{2010}]{kusano}  Kusano, T., Mari\'c, V., {\it Regularly varying solutions of perturbed Euler differential equations and related functional differential equation},  Publ. Inst. Math., 1 (2010), 88(102) 

\bibitem[\protect\citeauthoryear{Lehners, McFadden \& Turok}{2007}]{jean} Lehners J.-L., McFadden P., Turok N., {\it Colliding branes in heterotic M theory}, PhRvD, 75 (2007), 103510 
	
	
\bibitem[\protect\citeauthoryear{Lang}{2002}]{lang} Lang, S., {\it Algebra}, Springer (2002)

\bibitem[\protect\citeauthoryear{Liddle \& Lyth}{2000}]{liddle} Liddle A.~R., Lyth D.~H., {\it Cosmological Inflation and Large-Scale Structure}, cils.book, 414 (2000)

\bibitem[\protect\citeauthoryear{Liddle}{2003}]{liddle2003} Liddle A., {\it An Introduction to Modern Cosmology}, imcs.book, 188 (2003)
	
\bibitem[\protect\citeauthoryear{Liddle \& Murdin}{2006}]{liddle2006} Liddle A., Murdin P., {\it The Cosmological Constant and its Interpretation},  EAA, Murdin P., ed. IOP Publishing Ltd  (2006)
	
\bibitem[\protect\citeauthoryear{Mari\'c \& Tomi\'c}{1990}]{maric1990} Mari\'c, V., Tomi\'c, M., {\it A classification of solutions of second order linear differential equations by means of regularly varying functions}, Publ. Inst. Math. (Belgrade),
58 (1990), 199

\bibitem[\protect\citeauthoryear{Mari\'c}{2000}]{maric} Mari\'c, V., {\it Regular Variation and Differential  Equations}, Springer, Berlin (2000)

\bibitem[\protect\citeauthoryear{Mijajlovi\'c., Pejovi\'c \& Ninkovi\'c}{2007}]{mijajlo2007} Mijajlovi\'c, \v Z., Pejovi\'c, N. \& Ninkovi\'c, S., {\it Nonstandard Representations of Processes in Dynamical Systems}, AIP Conf. Proc, 934 (2007), 151

\bibitem[\protect\citeauthoryear{Mijajlovi\'c et al.}{2012}]{mijajlo2012} Mijajlovi\'c, \v Z., Pejovi\'c, N., \v Segan, S.,  Damljanovi\'c, G., {\it On asymptotic solutions of Friedmann equations}, Appl. Math and Computation, 219 (2012), 1273--1286

\bibitem[\protect\citeauthoryear{Mijajlovi\'c, Pejovi\'c \& Mari\'c}{2015}]{mijajlo2015} Mijajlovi\'c \v Z., Pejovi\'c N., Mari\'c V., {\it On the $\varepsilon$ cosmological parameter}, SerAJ, 190 (2015), 25 

\bibitem[\protect\citeauthoryear{Molchanov, Surgailis \& Woyczynski}{1997}]{molchanov} Molchanov S.A., Surgailis D., Woyczynski, W.A., {\it The large-scale structure of the universe and quasi-Voronoi tessalation of schock fronts in forced Burgers turbulence in $R^d$}, Ann. Appl. Probability, 7 (1997), 200
	
	%\bibitem{narlikar} Narlikar, J.V., An Introduction to  Cosmology. Cambridge Univ. Press, Cambridge (2002).
	
	%     \bibitem{16} Omey, E., Regular variation and its application to second order linear differential equation,
	%           Bull. Soc. Math. Belg. 32, 207 (1981)
	%     \bibitem{16} Peacock, J.A., Cosmological
	%           Physics, 682, Cambridge Univ. press, Cambridge (1999)

\bibitem[\protect\citeauthoryear{Peebles}{2017}]{peeble} Peebles P.~J.~E., {\it Growth of the nonbaryonic dark matter theory}, NatAs, 1 (2017), 0057 

	

\bibitem[\protect\citeauthoryear{Planck Collaboration et al.}{2016}]{planck} Planck Collaboration, et al., {\it Planck 2015 results. XIII. Cosmological parameters}, A\&A, 594 (2016), A13 


\bibitem[\protect\citeauthoryear{Perlmutter et al.}{1995}]{perlmutter} Perlmutter S., et al., {\it A supernova at Z = 0.458 and implications for measuring the cosmological deceleration}, ApJ, 440 (1995), L41 


\bibitem[\protect\citeauthoryear{Riess et al.}{2004}]{riess} Riess A.~G., et al., {\it Type Ia Supernova Discoveries at z > 1 from the Hubble Space Telescope: Evidence for Past Deceleration and Constraints on Dark Energy Evolution}, ApJ, 607 (2004), 665 


\bibitem[\protect\citeauthoryear{Schmidt et al.}{1998}]{schmidt} Schmidt B.~P., et al., {\it The High-Z Supernova Search: Measuring Cosmic Deceleration and Global Curvature of the Universe Using Type IA Supernovae}, ApJ, 507 (1998), 46 




\bibitem[\protect\citeauthoryear{Seneta}{1976}]{seneta} Seneta E., {\it Regularly varying functions}, Springer, Berlin (1976)

\bibitem[\protect\citeauthoryear{Spergel et al.}{2003}]{spergel} Spergel D.~N., et al., {\it First-Year Wilkinson Microwave Anisotropy Probe (WMAP) Observations: Determination of Cosmological Parameters}, ApJS, 148 (2003), 175 
	
\bibitem[\protect\citeauthoryear{Stern}{1997}]{stern} Stern I., {\it The Effect of Lacunarity on
        the Convergence of Algorithms for Scaling Exponents}, ASPC, 125 (1997), 222 


	%\bibitem{stroyan} Stroyan, K.D., Luxemburg, W.A.J.: Introduction to the theory of infinitesimals,  Academic Press, NY (1976).
	
	%\bibitem{swart}  de Swart, J. G., Bertone, G. and van Dongen, J. How dark matter came to matter. Nat. Astron. 1, 0059 (2017).

\bibitem[\protect\citeauthoryear{Tegmark et al.}{2004}]{tegmark} Tegmark M., et al., {\it  Cosmological parameters from SDSS and WMAP}, PhRvD, 69 (2004), 103501 



% Please avoid comments such as "For a review'', "For some examples",
% "and references therein" or move them in the text. In general,
% please leave only references in the bibliography and move all
% accessory text in footnotes.

% Also, please have only one work for each \bibitem.


\end{thebibliography}
\end{document}